\documentclass[final,12pt,a4paper]{elsarticle}

\usepackage{amssymb}
\usepackage{hyperref}
\usepackage{xcolor} 

\newcommand{\nivel}[0]{\textsc{Nivel}}
\newcommand{\niveltwo}[0]{\textsc{Nivel}${}^2$}
\newcommand{\Undulate}[0]{\textsc{Undulate}}

\journal{arXiv}

\begin{document}

\begin{frontmatter}



\title{nivel2: A web-based multi-level modelling environment built on a relational database}



\author[inst1]{Timo Asikainen\corref{cor1}}
\ead{timo.o.asikainen@helsinki.fi}
\author[inst1]{Tomi Männistö}
\ead{tomi.mannisto@helsinki.fi}
\author[inst1]{Eetu Huovila}
\ead{eetu.huovila@helsinki.fi}
\cortext[cor1]{Corresponding author.}

\affiliation[inst1]{organization={University of Helsinki, Department of Computer Science}}

\begin{abstract}
We introduce the \textsf{nivel2} software for multi-level modelling. Multi-level modelling is a modelling paradigm where a model element may be simultaneously a type for and an instance of other elements under some constraints. This contrasts traditional modelling methods, such as the UML, where an element may not be a class and an object simultaneously. In contrast with previous approaches to multi-level modelling, the \textsf{nivel2} software utilises an industrial scale relational database for data storage and reasoning. Further, a web-based user interface is provided for viewing and editing models. The architecture enables multiple users in different roles working on the same models at various levels of abstraction at the same time.
\end{abstract}


\end{frontmatter}


\section{Motivation and signifigance}
\label{sec:motivation}

Historically, multi-level modelling has been seen as an extension of two-level modelling, where the main modelling element has been \emph{class} and the instances of \emph{objects} of the classes have comprised the second, often implicit, model layer; see e.g. \cite{Atkinson2008,Atkinson2015}. The predominant modelling paradigm has been graphical modelling tools purported for creating class diagrams, and a number of tools based on the paradigm have been suggested, see, e.g. \cite{Guerin2022,Balaban2022,Rodriguez2021}. 
The context for such modelling has typically been software engineering, and the primary focus of modelling has been the design and, in some cases, the generation of software. The resulting models have typically been assumed to stored in a single file or a set of files, the essential feature of which being that a model can be processed in memory and edited using a software engineering tool, such as an Eclipse plugin.

However, in practice, such an assumption may be overly restrictive. We believe that in order to be a feasible alternative to store enterprise data and to be accessible to different user groups, multi-level modelling must be designed to leverage the state-of-the-practice tools for storing and editing enterprise data. Towards this end, we will utilise relational database management systems (RDBMS) for storing and reasoning about the model data and browser-based applications, implemented using JavaScript extensions, for editing and viewing the data.

The software described in this paper is based on our previous work on multi-level modelling, the \nivel{} modelling language \cite{Asikainen2009}. So far, \nivel{} has had no relevant tool support. As for implementation, the modelling concepts of \nivel{} were given semantics by means of mapping to a formal language that could further be processed by the \emph{smodels} and \emph{lparse} tools, that have not been further developed for years.

\section{Conceptual extensions to multi-level modelling}
\label{sec:extensions}

This paper describes software implementing \niveltwo{}, a version of \nivel{}. Papers describing \niveltwo{} from a theoretical and conceptual point of view are still under preparation. There are a number of differences and extensions between \niveltwo{} and previous work on multi-level modelling. We will discuss these in the following subsections in brief.

\subsection{Class vs.\ clabject vs.\ entity}

The term \emph{class} used in \nivel{} is replaced by \emph{entity} in \niveltwo{}; the term commonly used in multi-level modelling approaches is \emph{clabject}.

\subsection{Association vs.\ reference}

Also, the \emph{association} modelling construct in \nivel{} has been replaced by \emph{reference} in \niveltwo{}. An association in \nivel{}, similar to associations in the Unified Modeling Language (UML)\footnote{See \url{https://www.uml.org/}}, could have two or more ends or \emph{roles}, each played by one or more classes. Associations could be instantiated similarly as classes along the levels. 
However, from a theoretical point of view, the association semantics are complex and overlap with the class semantics. Therefore, in \niveltwo{}, the notion of \emph{reference} is adopted, with the abstract syntax that an entity may have a named reference to other entities termed the \emph{targets}. When the entity with a reference is instantiated, the reference targets are likewise instantiated: the reference targets in the type are \emph{possible types} for the reference targets in the instance. Consequently, an entity with its reference may be considered an association, with the number of references as its arity.

\subsection{Levels and potencies}

In \nivel{}, we adopted a strict notion of levels: each class was on a specific levels, and other relations than \emph{instance-of} between classes on different levels were not allowed. However, in recent research on multi-level modelling, arguments for more relaxed notions of levels haven been made, see e.g. \cite{Kuhne2022}. We follow the trend and drop the notion of strict metamodelling in \niveltwo{}. However, we still require, for example, that the transitive closure of the \emph{instance-of} relation is asymmetric.

As for potency, we retain the notion of potency for attributes in \nivel{}: an attribute with potency $p$ in class/entity $e$ must become a value in an instance of order $p$ of $e$. Also, we introduce a similar notion of potency for references. Let $e$ be an entity that has a reference named $r$ with potency $p$ with set of targets $T = \{t_1, \ldots, t_n\}$. Given an entity $i$ of type $e$, $i$ has reference named $r$ with potency $p-1$ and each target $s$ is an instance of some $t_i \in T$. A reference may not have potency $p \leq 0$.

A more elaborate discussion on various notions of potency will be provided in the subsequent publication on \niveltwo{}.

\subsection{Functions of entities}

Finally, the \emph{functions} are defined in \niveltwo{} as instances of a special \niveltwo{entity} called \textsf{function}. A type $t$ may refer to an instance of \textsf{function}, and the function can be \emph{run} on an instance of $t$.

A function has a structure defined in terms of other functions and \emph{actions}: these are collectively defined as \emph{steps} of the function. The steps are represented as an ordered reference called \textsf{steps} of \textsf{function}, with the functions and actions as its targets. When run, the actions of the function form the payload of the function. In the current implementation, actions are mapped to calls of HTTP API endpoints with input and output data; the type of endpoint (e.g. GET, POST) and the address are defined as attributes of the action. 



\section{Software description}

In this section, we describe the \textsf{nivel2} software: first, its technical architecture and thereafter, its main functionalities.

\subsection{Software architecture}
The core elements of the \niveltwo{} software are a \emph{relational database} and a \emph{data interface} called \textsf{datapi} for interacting with the database, and a user interface (\textsf{ui}) intended for users to interact with \niveltwo{} entities.

\subsubsection{Database}

The relational database was developed using the Azure SQL Database cloud service\footnote{See \url{https://azure.microsoft.com/en-us/products/azure-sql/database/}} by Microsoft Corporation, which implements major parts of the SQL standards.

\begin{figure*}
    \centering
    \includegraphics[width=\textwidth]{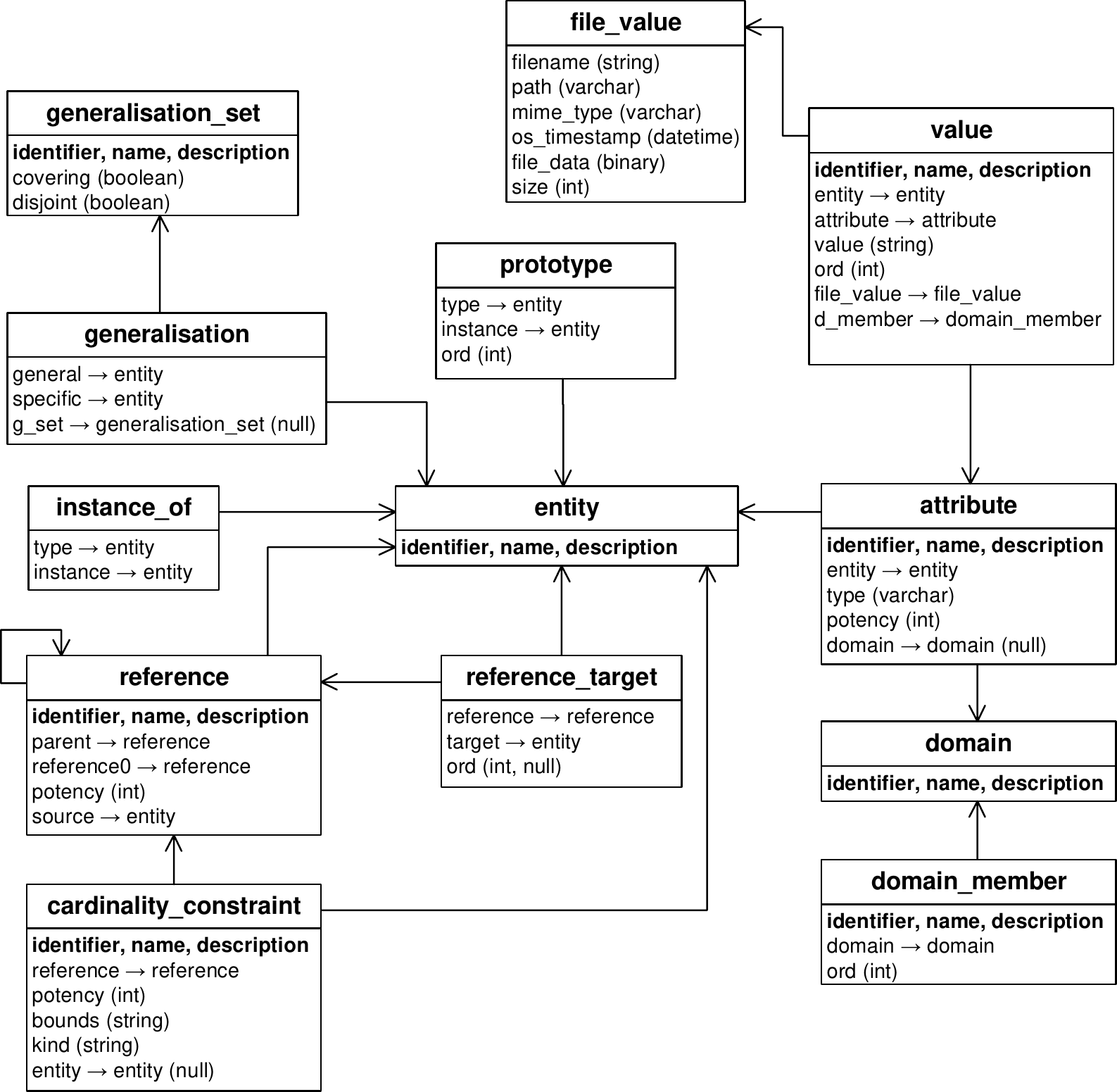}
    \caption{Database tables used to store \niveltwo{} data. The tables constitute the \textsf{nivel} schema in the database, the name of which may be changed based on the installation (the name \textsf{data} was used in development). Legend: Each table has an automatically generated identity value (integer) named id as its primary key. The single line ``\textbf{identifier, name, description}'' refers to three distinct columns, each of string (varchar) type. A column referencing another table is denoted with an arrow after the column name, followed by the name of the target table; each referenced column is the id value of the referenced table. In addition, the references are visualised graphically using an arrow from the referencing table pointing to the referenced table.}
    \label{fig:database}
\end{figure*}

The database tables are illustrated in Figure~\ref{fig:database}. As is evident from the table definitions, the data in the database is stored as individual facts: data related to an entity is spread across multiple tuples in multiple tables. This is, in contrast, to object  databases and, e.g.\ JSON (JavaScript Object Notation), where data related to a single object or entity is typically saved in one place. The motivation for the selected schema is twofold: first, the style is natural in a relational database; second, reasoning about the data is more straightforward when each kind of fact is stored in isolation from other kinds of facts.

Included in the database are stored procedures for querying the data for various purposes. The procedures are implemented using the T-SQL language specific to the Microsoft SQL Server family of products. In short, given an entity and usage (edit, instantiate, generalise), the database is queried for data related to the entity and a facet of the data relevant to the usage is returned. Most importantly, an entity has different type (when instantiated) and object (when edited or viewed) facets. Different usages are discussed in more detail in Section~\ref{subsec:functionalities}.

\subsubsection{Data interface -- \textsf{datapi}}
\label{sec:datapi}

The database can be accessed using an API component \emph{datapi} implemented in Python\footnote{See \url{https://www.python.org/}}. The emp application framework is used to enable  access to the implemented services. In the production setting, the Flask application is run using the Nginx web server\footnote{See \url{https://www.nginx.com/}}. In practical terms, the services can be accessed through the HTTP (Hyper-Text Transfer Protocol) using a number of endpoints.

The \textsf{datapi} components operate mainly on the data returned by the stored procedures in the database. The data returned by the procedures is still essentially in a table form, with data related to different entities gathered together. The endpoint implementations convert the data into entity form, where data related to an entity is stored in a single place. The entity form is more easily accessible to, e.g., the user interface.

The component also provides an endpoint for that can be used to run \niveltwo{} functions. 

\subsubsection{User interface -- \textsf{ui}}
\label{sec:uicomponent}

A generic interface component, implemented in React\footnote{See \url{https://reactjs.org/}}, a JavaScript library, can be run in a browser and used to edit, view, instantiate and specialise \niveltwo{} entities. Also, the interface enables running functions on entities and viewing the execution results.

\subsubsection{Scheduled execution of functions -- \textsf{observer}}

The \textsf{observer} component is intended for running scheduled tasks. The component is implemented in Python and interacts with both directly with the database 
as well as the \textsf{datapi} component in order to run functions; see the description of the functionality below.

\subsubsection{Gathering logs and other usage data from services -- \textsf{log\_data\_api}}

The \textsf{log\_data\_api} component is implemented in Python, with Flask and nginx used to provide an API similarly as for \textsf{builder}, see Section~\ref{sec:datapi}. The component stores the data it receives in the database, to which it has an ODBC connection. When the log data stemming from different services is stored in a centralised database, log data can easily be accessed for debugging and other forms of analysis.

\subsubsection{Converting \niveltwo{} data to different text formats -- \textsf{converter}}

The component is implemented in Python, with Flask and nginx used to provide an API similarly as for \textsf{builder}, see Section~\ref{sec:builder}. As its name suggests, the component can be used to perform \emph{conversions} from one form to another. Such a conversion is defined as a \niveltwo{} function on an entity. The function contains a set of \emph{patterns}, one of which is the \emph{root pattern}. A pattern may contain \emph{data placeholders} that refer to the data being converted: during the conversion, the placeholders are substituted with data from the entity. In the conversion process, the patterns are recursively applied to the entity data, starting from the root pattern. 

\subsection{Software functionalities}
\label{subsec:functionalities}

\subsubsection{User interface for multi-level models}

\begin{figure*}
    \centering
    \includegraphics[width=1.35\textwidth]{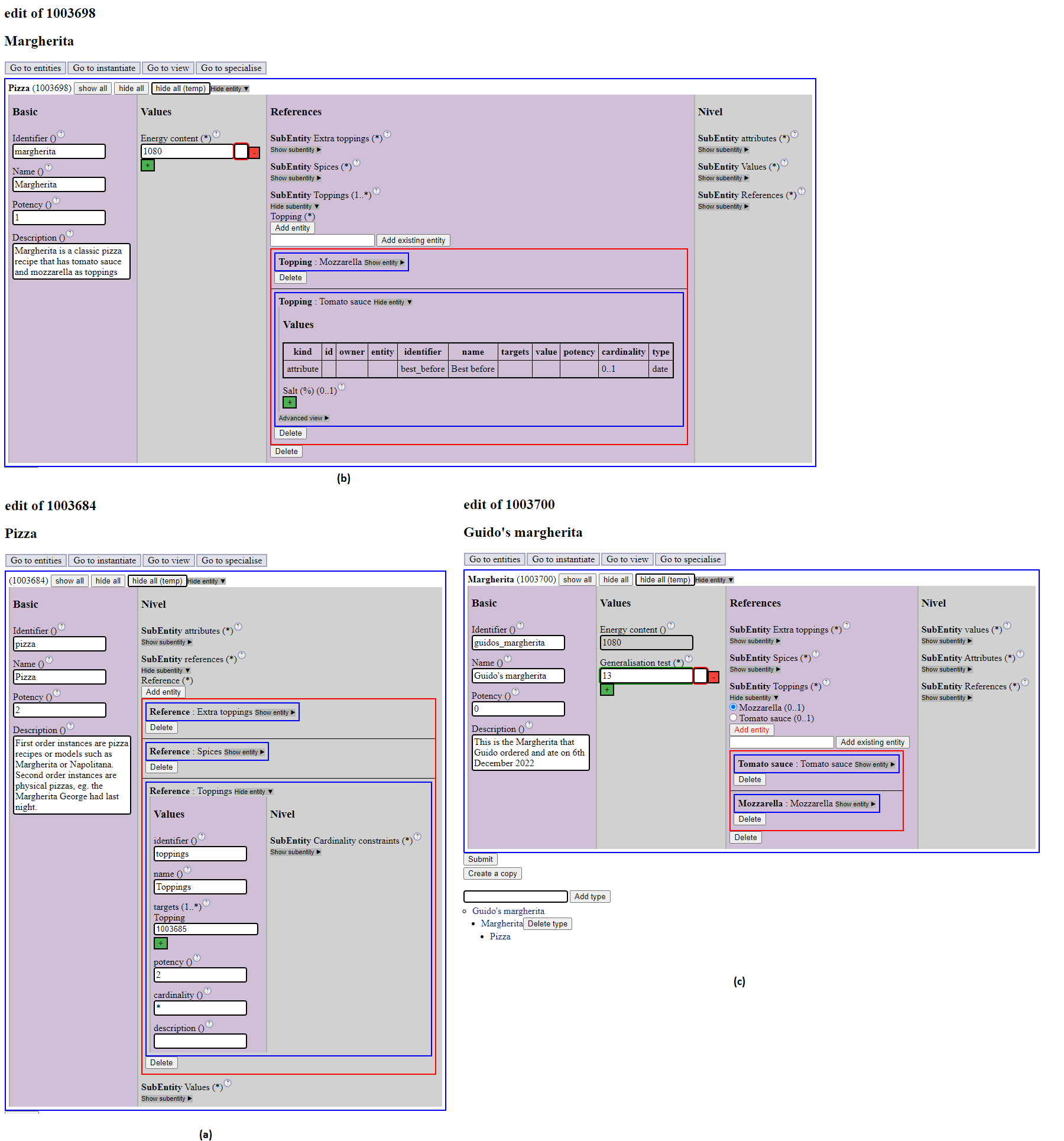}
    \caption{The \niveltwo{} user interface, illustrated with pizzas at various levels of abstraction. See the main text for descriptions of the panels (a)--(c).}
    \label{fig:ui}
\end{figure*}

The user interface is illustrated in Figure~\ref{fig:ui}. The  panel~(a) illustrates the basic entity \textsf{Pizza}, the instances of which are pizza recipes, such as Margherita. The pizza is given an identifier, a name and a description (on the left) and defined a reference to toppings (on the right, with entity id 1003685 being the corresponding root type for toppings). Further, panel~(b) illustrates an instance of \textsf{Pizza}, namely \textsf{Margherita}, a classic pizza recipe. Shown in the edit mode, the \textsf{energy content} (defined for \textsf{Pizza} but not shown in the panel for the lack of space) can be given a value in this context. Further, the panel shows the familiar toppings of Margherita, namely mozzarella and tomato sauce: these are shown under the \textsf{toppings} reference as targets, namely the \textsf{Mozzarella} and \textsf{Tomato sauce} entities, both instances of \textsf{Topping}.

Finally, panel~(c) illustrates an entity, \textsf{Guido's margherita}, representing a physical pizza. This entity could still be modified through, e.g. omitting either of the toppings. In addition, there are additional references (\textsf{extra toppings}) and \textsf{spices}) that are defined for \textsf{Pizza} but are not shown or discussed due to lack of space in detail. However, in general terms, the two references illustrate the idea that spices, such as garlic and oregano, can be included in any pizza. The same applies to extra toppings that can be included in a pizza (typically at cost) from a list of toppings (all instances of \textsf{Topping}). The two references, equipped with suitable \emph{cardinality constraints}, are used to model these aspects of pizzas.

\subsubsection{Scheduled execution of functions -- \textsf{observer}}

The component runs in two stages.

In the first stage, the observer runs a command in the database to fetch a list of \emph{observation targets}. Each observation target includes the following data:

\begin{itemize}
    \item[--] \textsf{query}, expressed in SQL, to be executed against the database
    \item[--] \textsf{interval} time (in seconds) that specifies how often the observation target is to be run
    \item[--] \textsf{function} and \textsf{parent\_reference} that identify a \textsf{Nivel2} function to be executed in relation to the observation target
\end{itemize}

For each observation target (as defined above), the following is done:

\begin{enumerate}
    \item The \textsf{query} is executed, and for each row returned by the query:
    \item The function defined by the \textsf{function} and \textsf{parent\_reference} (see above) is run using \textsf{datapi} for the entity defined by the mandatory *entity* column in the row in point 1
\end{enumerate}

As an example, an observation target could be to run maintenance tasks on all instances of an entity (type) that match a certain condition. For example, permanently delete all emails classified as junk at least 30 days ago. 




\section{Impact overview}
\label{sec:impact}

So far, we have applied the software in our own research in the xCESE (eXtreme Continuous Experimentation in Software Engineering) research project with success. More specifically, we have applied the software in order to systematically represent software experimentation knowledge at multiple levels of abstraction. Our work is described in more detail from a theoretical point of view in \cite{Asikainen2022}, and its implementation, \textsf{Undulate} software is available as open source\footnote{See \url{https://version.helsinki.fi/xcese_public/undulate}}. In short, \Undulate{} demonstrates that knowledge related to continuous experimentation can be represented using \niveltwo{}, which is demonstrated by the fact that the \niveltwo{} representation can be used to run experiments and accumulate their results automatically.

The levels of abstraction in continuous experimentation are:

\begin{enumerate}
    \item Basic concepts related to an experiment, such as \textsf{Experiment} and \textsf{Group} (either control or test) 
    \item Instances of the basic concepts that represent the design of an experiment, e.g. testing different link colours (\textsf{ColourExperiment}), where the groups would be identified by a different colour each
    \item Execution data of the experiment, e.g. runs of experiment design with concrete start times and users assigned to the groups
\end{enumerate}

We believe that representing experimentation data at all levels of abstraction using a multi-level modelling language such as \emph{nivel2} adds rigour to work and thus makes the work more systematic. Once the basic concepts are defined, the \textsf{nivel2} software makes it easier to adhere to the predefined structure of concepts. Also, the \textsf{nivel2} interface can be used to navigate from experiment results (Level 3) to their design (Level 2). Without uniform data representation, data at Level~1 and possibly even Level~2 would be represented using a programming language or, alternatively, configuration files without consistency checks. Therefore, we believe that resorting to multi-level modelling reduces the amount of coding required and makes the process, in this case, experimentation in software engineering, more accessible.

This publication marks the release of the \textsf{nivel2} software to the public. However, there are plans to apply the software in various research projects within our research group. Also, making the code publicly available may attract a larger audience, as well as publishing the theoretical papers on \textsf{nivel2} that are still work in progress. Also, provided that sufficient funding can be arranged, the centralised database enables a public \textsf{nivel2} server to be hosted, thus supporting easy access and experimenting on \niveltwo{} concepts to both the scientific and business communities.


\section{Acknowledgements}
\label{sec:acknoledgements}

The work was supported by the Academy of Finland (project 317657). 



\bibliographystyle{elsarticle-num} 
\bibliography{xcese}

\section*{Required Metadata}
\label{metadata}

\section*{Current code version}
\label{version}

Ancillary data table required for subversion of the codebase. Kindly replace examples in right column with the correct information about your current code, and leave the left column as it is.

\begin{table*}[!ht]
\begin{tabular}{|l|p{6.5cm}|p{6.5cm}|}
\hline
\textbf{Nr.} & \textbf{Code metadata description} & \textbf{Please fill in this column} \\
\hline
C1 & Current code version & v1 \\
\hline
C2 & Permanent link to code/repository used for this code version & \url{https://version.helsinki.fi/xcese_public/nivel2} \\
\hline
C3  & Permanent link to Reproducible Capsule & \\
\hline
C4 & Legal Code License  & MIT \\
\hline
C5 & Code versioning system used & git \\
\hline
C6 & Software code languages, tools, and services used & Python, JavaScript, React, nginx, Microsoft SQL Server, Docker \\
\hline
C7 & Compilation requirements, operating environments \& dependencies & Can be run on Linux with the included \texttt{Dockerfile}'s \\
\hline
C8 & If available Link to developer documentation/manual & -- \\
\hline
C9 & Support email for questions & timo.o.asikainen@helsinki.fi \\
\hline
\end{tabular}
\caption{Code metadata (mandatory)}
\label{cmdm} 
\end{table*}




\end{document}